\documentclass[twocolumn,showpacs,preprintnumbers]{revtex4}

\usepackage{amsmath,bbm,epsfig}

\newcommand{\be}{\begin{equation}}
\newcommand{\ee}{\end{equation}}
\newcommand{\ba}{\begin{array}}
\newcommand{\ea}{\end{array}}
\newcommand{\bqa}{\begin{eqnarray}}
\newcommand{\eqa}{\end{eqnarray}}

\newcommand{\tr}{\mbox{Tr}}
\newcommand{\bra}[1]{\ensuremath{\langle #1 |}}
\newcommand{\ket}[1]{\ensuremath{| #1 \rangle}}
\newcommand{\prj}[1]{\ensuremath{| #1 \rangle \langle #1 |}}
\newcommand{\ovl}[2]{\ensuremath{\langle #1 | #2 \rangle}}

\begin{document}

\title{Concurrence of mixed multi-partite quantum states}

\author{Florian Mintert$^{1,2}$, Marek Ku\'s$^2$ and Andreas Buchleitner$^1$}

\affiliation{
  $^1$Max-Planck-Institut f\"ur Physik komplexer Systeme,
  N\"othnitzerstr. 38, D-01187 Dresden
}
\affiliation{
  $^2$Centrum Fizyki Teoretycznej Polskiej Akademii Nauk,
  Aleja Lotnik\'ow 32/46,
  PL-02-668 Warszawa
}

\date{\today}
\begin{abstract}
We propose generalizations of concurrence for multi-partite quantum
systems that can distinguish qualitatively distinct quantum correlations.
All introduced quantities can be evaluated efficiently for arbitrary
mixed sates.
\end{abstract}

\pacs{03.67.-a, 03.67.Mn, 89.70.+c}

\maketitle

Quantum correlations display one of the crucial - and arguably the least
understood - qualitative differences between classical and quantum systems.
While we start to develop some kind of intuition for nonclassical
correlations which arise in quantum systems composed of two subsystems, our
comprehension remains rather strained when we deal with a larger number of
constituents.
This qualitative difference between bi- and multi-partite quantum systems
originates - among others - from the fact that in
bipartite quantum systems there are no qualitatively different quantum
correlations, {\it i.e.} any state $\varrho$ can be prepared with
local operations and classical communication (locc), starting out from a
maximally entangled state.
Therefore, the entanglement of bipartite states can be well characterised by
a single scalar quantity, such as, {\it e.g.}, the entanglement of formation
\cite{eof}.
In multi-partite systems this is no more true. For example,
the label `maximally entangled' can be justified for both
GHZ- \cite{ghz} and W-states, though none of them can be prepared
from the other using only locc \cite{duer00}.
Hence, they are characterized by qualitatively different, inequivalent quantum
correlations.

Also the very definition of multi-partite separability and entanglement
requires some refinement as compared to bipartite systems:
An $N$-partite system is described by a Hilbert space $\cal H$ that decomposes
into a direct product of $N$ subspaces
${\cal H}={\cal H}_1\otimes\hdots\otimes{\cal H}_N$,
where the dimension of the space ${\cal H}_i$ will be denoted by $n_i$.
A multi-partite state acting on $\cal H$ is separable \cite{wern89}, if
it can be
written as a convex sum of direct products of sub-system states
\be
\varrho=\sum_ip_i\ \varrho_1^{(i)}\otimes\hdots\otimes\varrho_N^{(i)}=
\sum_ip_i\bigotimes_{j=1}^{N}\varrho_j^{(i)}\ ,\
p_i>0\ .
\ee
In such a state all correlations between any of the subsystems are
of classical nature, and can be described in terms of the classical probabilities
$p_i$.
However, the tensorial structure of ${\cal H}$ holds room for qualitatively
distinct quantum correlations which go beyond the classical framework.
A state is {\em bi-separable}, if it can be written as a convex sum of states
which each decompose into a direct product of at least two factors.
This  definition can be immediately generalized 
to {\em $m$-separability} ($m\le N$), what provides a fine graduation of
states according to their different degrees of separability.

In general, it is an open problem to decide on the degree of separability of
a given state $\varrho$.
Some substantial progress has been achieved with entanglement witnesses
\cite{witness02,witness04}, which
allow to distinguish some multi-partite entangled states from bi-separable ones.
Though, their use also has two severe drawbacks:
some a priori knowledge on the considered state is required, in order to
construct a suitable witness -- and there is no general prescription for such
construction.
Moreover, a witness can give reliable information only if it actually does
detect
the considered state -- if not, this can either be due to
the fact that the witness is not well-adapted for the given state, or
that the state simply does not contain the quantum correlations seeked for.
A secure answer is only obtained if all (infinitely many) different
witnesses are consulted.

In the present Letter, we follow an alternative route, by formulating a
general recipe for the characterization of targeted separability properties of
arbitrary mixed $N$-partite states. No a priori knowledge on the state under
scrutiny will be needed here. Our approach is built upon suitable
generalizations of the concurrence of bipartite quantum states, and on
recently derived \cite{lb}, in general tight \cite{flodiss} lower bounds
thereof.

Let us start with a novel definition of pure state concurrence which is
different from the ones familiar from the published literature
\cite{wot98,run01}, though comprises these as special cases.
A simple definition on the level of pure states is an indispensable
prerequisite for the  treatment of mixed states, which we are finally aiming
at.

By analogy to the expectation value of an observable,
the expectation value of a suitably chosen linear, hermitean operator comes to
mind as a simple choice.
However, there is none such that all expectation values with respect to
entangled states are strictly positive, whereas they vanish for
all separable states - so that entanglement can be identified unambigously.
Though, we will see that the expectation value of linear, hermitean operators
$\cal A$ with respect to {\em two} copies of a pure state can
capture entanglement properties very well, what provides a solid
basis for generalizations for multi-partite mixed states.

\begin{figure}[t]
\epsfig{file=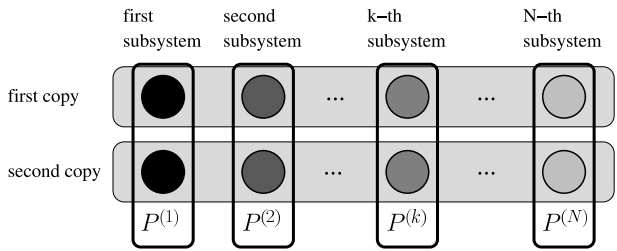,angle=0,width=0.45\textwidth}
\caption{
The concurrence of a pure state $\ket{\Psi}$ is defined in terms of two copies
(shown schematically as grey boxes)
of $\ket{\Psi}$, and of an operator $\cal A$ acting on them.
$\cal A$ is composed of projectors $P^{(i)}$ acting on the two copies of
subsystem $i$ only (shown as circles in different grey-scales).
\label{fig1}}
\end{figure}

So which operators $\cal A$ serve our purpose?
Of course, concurrence needs to be invariant under local unitaries.
Since this must hold for any state, it is natural to
require that $\cal A$ itself bears this invariance property.
The projectors $P^{(i)}_-$ and $P^{(i)}_+$ onto the anti-symmetric and
symmetric subspaces
${\cal H}_i\wedge{\cal H}_i$  and ${\cal H}_i\odot{\cal H}_i$ of
${\cal H}_i\otimes{\cal H}_i$
\footnote{
The anti-symmetric (resp. symmetric) subspaces of
${\cal H}_i\otimes{\cal H}_i$ are spanned by the states that adopt a phase
shift of $\pi$ (resp. $0$) upon exchange of the two copies of
${\cal H}$.
Expressed in terms of some arbitrary, orthonormal basis
$\{\ket{\alpha_j}\}$
of
${\cal H}_i$, the projectors onto these spaces read
$P^{(i)}_\mp=\sum_{jk}
\left(\ket{\alpha_j\alpha_k}\mp\ket{\alpha_k\alpha_j}\right)
\left(\bra{\alpha_j\alpha_k}\mp\bra{\alpha_k\alpha_j}\right)/4
$.}
($i=1,\ldots N$ labels the individual subsystems)
exhibit the desired invariance property, since any local unitary transformation
${\cal U}_{i}\otimes{\cal U}_{i}$ on ${\cal H}_i\otimes{\cal H}_i$
commutes with the exchange of both copies.
Hence, they can be used as elementary building blocks of
${\cal A}=\bigotimes_{j=1}^N P_{s_j}^{(j)}$ ($s_j=\pm$),
which immediately inherits their invariance with respect to local unitaries.
Thus, we can define a generalized concurrence of a pure,
$N$-partite state $\ket{\Psi}$ as
\be
c(\Psi)=\sqrt{\bra\Psi\otimes\bra\Psi{\cal A}\ket\Psi\otimes\ket\Psi}\
,
\label{conc_pure}
\ee
where
${\cal A}$
acts on $\bigotimes_{i=1}^N{\cal H}_i\otimes{\cal H}_i$
\footnote{Note that $\cal A$ is acting on
$\bigotimes_{i=1}^N{\cal H}_i\otimes{\cal H}_i$,
whereas
$\ket{\Psi}\otimes\ket{\Psi}\in
(\bigotimes_{i=1}^N{\cal H}_i)\otimes
(\bigotimes_{i=1}^N{\cal H}_i)$.
Though, the two different spaces are isomorphic, and it is
obvious how to identify their corresponding elements.},
as shown schematically in fig.~\ref{fig1}.

The above expression for ${\cal A}$ in terms of products of $P^{(i)}_\pm$
allows to tailor eq.~(\ref{conc_pure}) such as to address specific types of
correlations, as follows from inspection of the action of $P^{(i)}_\pm$ on
a two-fold copy
$\ket{\xi^{(i)}}\otimes\ket{\xi^{(i)}}\in
{\cal H}_i\otimes{\cal H}_i$
of a one-party state:
Since any such twofold copy is symmetric, the expectation value of $P_-^{(i)}$
vanishes identically, whereas the corresponding expression for $P_+^{(i)}$
gives unity.
Now, consider an $N$--partite state that separates into a one-party state
(let's take subsystem $N$, for simplicity) and a state of the remaining $N-1$
subsystems.
If ${\cal A}$ in eq.~(\ref{conc_pure}) comprises the term $P_-^{(N)}$,
the corresponding concurrence necessarily vanishes, thus highlighting
the vanishing entanglement between subsystem $N$ and the rest.
If, instead, one choses to incorporate the term $P_+^{(N)}$ in
eq.~(\ref{conc_pure}),
the respective concurrence will be sensitive to $(N-1)$--partite
correlations between the $N-1$ first subsystems.
This argument can be iterated recursively down to bi-partite correlations,
such that the original concurrence \cite{wot98,run01} of bipartite systems
emerges with the specific choice
${\cal A}=4 P_-^{(1)}\otimes P_-^{(2)}$.

Similarly to the case of bipartite systems, also our presently introduced
$N$-partite concurrences can be generalised for mixed states via
their convex roofs,
\be
c(\varrho)=\inf\sum_ic(\psi_i)\ ,
\label{roof1}
\ee
where the infimum is to be taken among all sets of properly normalized
$N$-partite states
such that $\varrho=\sum_i\prj{\psi_i}$.
Given one such set - {\it e.g.}, the spectral decomposition $\{\ket{\phi_i}\}$
of $\varrho$ -
all legitimate decompositions can be constructed as linear combinations
$\ket{\psi_i}=\sum_j V_{ij}\ket{\phi_j}$,
with the additional constraint that the complex prefactors $V_{ij}$ define a
left unitarity matrix, {\it i.e.}
$\sum_i V_{ij}^\ast V_{ik}=\delta_{jk}$ \cite{schroed}.
The above optimization over decompositions into pure states is thus
equivalent to an optimization over left unitary tranformations,
acting on the tensor $\hat{\cal A}$, with elements
$\hat{\cal A}_{jk}^{lm}=\bra{\phi_j}\otimes\bra{\phi_k}
{\cal A}\ket{\phi_l}\otimes\ket{\phi_m}$
\cite{bad02,lb}.

Since left unitary transformations define a high dimensional, continuous set,
the explicit evaluation of eq.~(\ref{roof1}) tends to be rather
cumbersome for a general mixed state.
However, techniques which were devised to ease that task in the
bipartite case \cite{lb} can be generalized in a straightforward manner,
because the algebraic structure of the above
$N$-partite concurrences (\ref{conc_pure}) is strictly identical to the
bipartite definition:
$\cal A$ and therefore also $\hat{\cal A}$ is hermitean and positive,
what allows the decomposition
$\hat{\cal A}_{jk}^{lm}=\sum_\alpha
T^\alpha_{jk}\left(T^\alpha_{lm}\right)^\ast$,
in terms of the matrices
$T_{jk}^\alpha=\bra{\phi_j}\otimes\ovl{\phi_k}{\chi^\alpha}$, which
in turn are defined via the spectral decomposition
${\cal A}=\sum_\alpha\prj{\chi_\alpha}$.
Analogous to \cite{lb} one can invoke the Cauchy-Schwarz- and the triangle
inequality and bound the concurrence of an arbitrary mixed state from below by
\be
c(\varrho)=\inf_V\sum_i\sqrt{
\sum_\alpha\left|\left[VT^\alpha V^T\right]_{ii}\right|^2}\ge
\inf_V\sum_i\left|\left[V\tau V^T\right]_{ii}\right|\ ,
\label{bound}
\ee
with
$\tau=\sum_\alpha z_\alpha T^\alpha$, and the inequality holds for an
arbitrary set of complex numbers $z_\alpha$, such that
$\sum_\alpha|z_\alpha|^2=1$ \cite{lb}.

Without loss of generality, we can assume the number of
factors $P_-^{(i)}$ in the explicit representation of $\cal A$ to be even,
since otherwise the concurrence vanishes identically
\footnote{
Any choice with an odd number of factors $P_-^{(i)}$ is anti-symmetric with
respect to an exchange of both copies of ${\cal H}$ (the entire system, and
not a single subsystem!), such that the expectation value with respect to the
twofold copy $\ket{\Psi}\otimes\ket{\Psi}$ of a state ({\it i.e.}, a symmetric
object) vanishes.}.
Therefore, any $\ket{\chi_\alpha}$ is symmetric with respect to the exchange
of the two copies of ${\cal H}$, and thus, the matrices $T^\alpha$ are complex
symmetric.
Consequently, the infimum on the right hand side of eq.~(\ref{bound}) can be evaluated
algebraically \cite{wot98,uhl00},
and the explicit solution in terms of the singular values $\lambda_j$
(labeled in decreasing order) of $\tau$
reads $\lambda_1-\sum_{j>1}\lambda_j$.
The choice of the prefactors $z_\alpha$ -- that determine $\tau$ -- can be
optimized numerically in order to approach the optimal lower bound.
Furthermore, a purely algebraic, and in most cases excellent
\cite{qp,multipartdyn}
approximation for $c(\varrho)$ can be obtained by substituting $\tau$ by the
matrix $\tau^{qp}$ with elements
${\tau^{qp}_{ij}=
\bra{\phi_1}\otimes\bra{\phi_1}{\cal A}\ket{\phi_i}\otimes\ket{\phi_j}/
\sqrt{\bra{\phi_1}\otimes\bra{\phi_1}{\cal A}\ket{\phi_1}\otimes\ket{\phi_1}}}$.
Herein $\ket{\phi_1}$ is the eigenvector associated with the largest
eigenvalue in the spectral decomposition of $\varrho$.

The above does not only apply to the discrete set of
concurrences discussed so far, but also to the following
continous interpolation between them:
Instead of a single direct product of projectors onto symmetric and
anti-symmetric subspaces, one may equally well consider convex
combinations thereof,
\be
{\cal A}=\sum_{{\cal V}{\{s_i=\pm\}}\atop{\prod_{i=1}^Ns_i=+}} p_{\{s_i\}}
\bigotimes_{j=1}^N P_{s_j}^{(j)}\ ,\
p_{\{s_i\}}\ge 0\ ,
\label{defA}
\ee
where ${\cal V}{\{s_i=\pm\}}$ represents all possible variations of an
$N$-string of the symbols $+$ and $-$,
and the summation is restricted to contributions with an even number of
projectors onto anti-symmetric subspaces.
Hence, through the arbitrariness of the choice of the
$p_{\{s_i\}}$,
there is actually a continous family of concurrences, and we leave the
interpretation of such $N$-particle concurrences for arbitrary $p_{\{s_i\}}$
as an
open (and intriguing) question.

However, a discrete subset of $N$-partite
concurrences has a transparent physical interpretation, and precisely allows
to distinguish different categories of multi-partite entanglement.
As an illustration, let us focus on some examplary
tri- and four-partite concurrences, which target at specific types of
multi-partite quantum correlations (see also table~\ref{table}), in the
remainder of the present contribution.
Note that contenting ourselves with pure states, does not imply any
restriction,
since the concept of convex roofs guarantees that all properties on the level
of pure states immediately convey to mixed states,
as, {\it e.g.}, also utilized in \cite{multipartdyn}.

Bi-separability with respect to specific partitions of tri-partite states 
is easily detected by
$c_1^{(3)}$, with
${\cal A}=4P_+^{(1)}\otimes P_-^{(2)}\otimes P_-^{(3)}$
in eq.~(\ref{defA}) above, and,
analogously, by
$c_2^{(3)}$, with
${\cal A}=4P_-^{(1)}\otimes P_+^{(2)}\otimes P_-^{(3)}$,
and by $c_3^{(3)}$, with
${\cal A}=4P_-^{(1)}\otimes P_-^{(2)}\otimes P_+^{(3)}$.
Whereas $c^{(3)}_1$ and $c^{(3)}_2$ vanish identically for
bi-separable states like
$\ket{\psi}=\ket{\varphi^{(12)}}\otimes\ket{\zeta^{(3)}}$,
$c^{(3)}_3$ reduces to the bipartite concurrence of
$\ket{\varphi^{(12)}}$, {\it i.e.}, $c^{(3)}_3(\psi)=c(\varphi^{(12)})$.
Thanks to the above construction (\ref{roof1}) as convex roof, these
properties also pertain to mixed states.
Note, however, that $c^{(3)}_3(\varrho)$ for a
bi-separable mixed state $\varrho$ is {\em not} equivalent to
$c(\tr_3\varrho)$:
Whereas the latter expression gives the bipartite correlation of $\varrho$ only
for a product state $\varrho=\rho^{(12)}\otimes\sigma^{(3)}$,
$c^{(3)}_3$ is significantly more powerful, and performs the same task for any
state with arbitrary classical correlations between ${\cal H}_3$ and the
combined system of ${\cal H}_1$ and ${\cal H}_2$, {\it i.e.}, for states like
$\sum_ip_i\rho_i^{(12)}\otimes\sigma_i^{(3)}$, $p_i\geq 0$.
Arbitrary quantum correlations can be accounted for by
$C_3$ with $p_{+--}=p_{-+-}=p_{--+}=4$ and $p_{+++}=0$ in
eq.~(\ref{defA}), which vanishes only for
completely separable states.
For any bi-separable state its value is given by the
bipartite concurrence of the remaining entangled part.

Similarly, different degrees of separability are also captured in
larger systems.
For example, concurrences like $c^{(4)}_{ij}$, with
${\cal A}=16P_{s_1}^{(1)}\otimes P_{s_2}^{(2)}\otimes
P_{s_3}^{(3)}\otimes P_{s_4}^{(4)}$,
$s_i=s_j=+$, and $s_k=-$ for $i\neq k\neq j$,
determine with respect to which bipartite partition a mixed four-partite state
is separable, and reduce to the bi- resp. tri-partite concurrences of
the entangled remainder.
\begin{table*}
\begin{tabular}[t]{|l||l|l|l|l|||l||l|l|l|}\hline
$\ket{\psi}\in\otimes_{i=1}^3{\cal H}_i$ &
$c^{(3)}_1(\psi)$ & $c^{(3)}_2(\psi)$ & $c^{(3)}_3(\psi)$ & $C_3(\psi)$
& $\ket{\psi}\in\otimes_{i=1}^4{\cal H}_i$ &
$c^{(4)}_{12}(\psi)$ & $c^{(4)}_{34}(\psi)$ & $C_4(N)(\psi)$ \\ \hline
$\ket{\varphi^{(12)}}\otimes\ket{\zeta^{(3)}}$ &
$0$ & $0$ & $c(\varphi^{(12)})$ & $c(\varphi^{(12)})$ &
$\ket{\varphi^{(123)}}\otimes\ket{\zeta^{(4)}}$ &
$0$ & $2c^{(3)}_3(\varphi^{(123)})$ & $0$ \\ \hline
$\ket{\varphi^{(13)}}\otimes\ket{\zeta^{(2)}}$ &
$0$ & $c(\varphi^{(13)})$ & $0$ & $c(\varphi^{(13)})$ &
$\ket{\varphi^{(12)}}\otimes\ket{\zeta^{(34)}}$ & $c(\zeta^{(34)})\eta(\varphi^{(12)})$ &
$c(\varphi^{(12)})\eta(\zeta^{(34)})$ & $c(\varphi^{(12)})c(\zeta^{(34)})$ \\ \hline
$\ket{\zeta^{(1)}}\otimes$\ket{\varphi^{(23)}} &
$c(\varphi^{(23)})$ & $0$ & $0$ & $c(\varphi^{(23)})$ &
$\ket{\Psi_{GHZ}}$ & $ 2\sqrt{\sum_{i>j}\lambda_i\lambda_j}$ &
$2\sqrt{\sum_{i>j}\lambda_i\lambda_j}$ &
$2\sqrt{\sum_{i>j}\lambda_i\lambda_j}$ \\ \hline
\end{tabular}
\caption{
  Some examplary tri- and four-partite concurrences for
  bi-separable and GHZ-like states.
  For bi-separable states, the concurrences $c^{(3)}_i$ ($i=1,2,3$),
  $c^{(4)}_{12}$, and $c^{(4)}_{34}$ either vanish
  or reduce to the bi- resp.  tri-partite concurrence of the remaining
  entangled part
($\ket{\varphi^{(12)}}$, $\ket{\varphi^{(13)}}$, $\ket{\varphi^{(23)}}$,
  $\ket{\varphi^{(123)}}$, or $\ket{\zeta^{(34)}}$) of $\ket{\psi}$,
with $\eta(\phi)=\sqrt{1-c(\phi)^2/4}$.
  $C_4$ vanishes for all states where at least one particle is
uncorrelated with the other system components.
  \label{table}
}
\end{table*}

Moreover, there is also the concurrence $C_4$, defined through
${\cal A}=16P_-^{(1)}\otimes P_-^{(2)}\otimes P_-^{(3)}\otimes P_-^{(4)}$,
that characterizes separability properties independent of any pairing
of subsystems:
$C_4$ vanishes for any
state where at least one subsystem is uncorrelated with all other system
components.
In particular, for GHZ-like states
$\ket{\Psi_{\rm GHZ}}=\sum_i\sqrt{\lambda_i}\ket{iiii}$, $c_4$ yields
non-vanishing values, see table \ref{table}, whereas
it vanishes for $W$-states,
as it is in accord with observations that in tri-partite systems $W$-states
contain only bipartite correlations \cite{tangle}.
For two-level systems, $c^{(4)}$ can be used as a measure of
the usefulness of a given state for
multi-particle teleportation \cite{rigolin},
and, since ${\cal A}$ is of rank one, eq.~(\ref{bound}) does not only provide
a lower bound, but rather the exact concurrence of arbitrary mixed states.

As a last example, we would like to focus on the $N$-partite
generalization $C_N$ of $C_3$,
with $p_{\{s_i\}}=4$ for all $\{s_i\}$ in eq.~(\ref{defA}), except for
$p_{+\hdots +}=0$.
Defined for systems with an arbitrary number $N$
of subsystems,
$C_N$ can be shown \cite{monotone}
to be monotonously decreasing under {\it locc} operations,
such that it does not only allow to access separability properties, but also
is
an entanglement monotone \cite{vidal00}.
As already pointed out in \cite{multipartdyn},
$C_N$ can -- like any of the concurrences defined in
eq.~(\ref{conc_pure}) --
be expressed in terms of all reduced density matrices $\varrho_i$
\be
C_N(\Psi)=2^{1-\frac{N}{2}}
\sqrt{(2^N-2)\ovl{\Psi}{\Psi}^2- \sum_i\tr\varrho_i^2}\ ,
\label{cN}
\ee
where the multi-index $i$ runs over all $(2^N-2)$ subsets of the $N$
subsystems.

Like $C_3$, $C_N$ only vanishes for completely separable
$N$-partite states. Furthermore, it has the particularly nice property
that $C_N(\psi)$
reduces to $C_{N\mbox{-}1}(\xi)$ for any state $\ket{\psi}$ that
factorizes into a product state on one subsystem and on the
$(N-1)$-partite remainder $\ket{\xi}$.
This allows to compare the non-classical correlations inscribed in
multi-partite systems of variable size $N$.

In conclusion, we have seen that a discrete subset of the continuous family of
concurrences defined by eqs.~(\ref{conc_pure},\ref{defA}) allows for a
selective assessment of the separability properties of mixed multi-partite
quantum states. Given the lower bound (\ref{bound}), these quantities can be
evaluated efficiently \cite{qp,lb,multipartdyn}, and thus allow us, e.g., to
address important questions such as the time evolution and the scaling
properties (in terms of the system size) of entanglement in higher dimensional
quantum systems \cite{multipartdyn}
-- an objective which so far could only be accomplished for
the simple $2\times 2$ case \cite{eberly}.
Furthermore, our definition reveals a continuously parametrized identifier of
multipartite entanglement, an observation which still awaits its
physical/statistical interpretation.

We are indebted to Andr{\'e} Ricardo Ribeiro de Carvalho,
Rafa\l\ Demkowicz-Dobrza\'nski, and Karol \.Zyczkowski
for fruitful discussions, comments and remarks.
Financial support of VolkswagenStiftung  and the Polish Ministry of Science
through the grant No. PBZ-MIN-008/P03/2003 is greatfully acknowledged.

\bibliography{referenzen}

\end{document}